\documentclass[12pt]{article}

\usepackage{amssymb}
\usepackage[centertags]{amsmath}
\usepackage{rotating}
\usepackage[dvips]{color}	

\usepackage{epsf}
\usepackage{pazh_eng}

\tightenlines

\newcommand{\dg}{\mbox{$^\circ$}}
\def\fdg{\hbox{$.\!\!^\circ$}}
\sloppypar

\begin{document}

{\renewcommand{\thefootnote}{\fnsymbol{footnote}}

\title{\bf Segments of spiral arms of the Galaxy traced by classical Cepheids: Effects of age heterogeneity}

\author{\bf
Angelina V. Veselova, Igor' I. Nikiforov
}
\setcounter{footnote}{0}

\vspace{-0.9em}
\begin{center}

   \it{Department of Celestial Mechanics, Saint Petersburg State University, 
Universitetskij Prospekt 28, Staryj Peterhof, Saint Petersburg, 198504
Russia; {\it linav93@yandex.ru (AVV)}\\ }

\end{center}

\vspace{-0.3em}

\sloppypar 
\vspace{2mm}
\noindent
{\bf Abstract}---{We investigated the dependence} of the parameters of the segments of spiral arms of the
Galaxy on the age of classical Cepheids. The catalog of Cepheids (Mel'nik et al. 2015)
was divided into two samples---relatively young ($P>9^\text{d}$) and relatively old
($P<9^\text{d}$) objects. The parameters of the spiral structure were determined both for
two samples separately and jointly for the combination of two systems of segments traced
by young and old objects. For most of the segments, their parameters for young and old
objects differ significantly.  Taking into account the difference between the two segment
systems, we obtained the estimate $R_0$ equal to $7.23^{+0.19}_{-0.18}$\,kpc, which in the
modern LMC calibration corresponds to the value of
$R_0={8.08^{+0.21}_{-0.20}}|_{\text{stat.}} {}^{+0.38}_{-0.36}|_{\text{cal.}}$\,kpc.  It
is shown that the displacement between the segments is not reduced to the effect of
differential rotation only. To interpret this displacement for objects of Perseus and
Sagittarius-2 segments  we carried out a dynamic
modeling of the change in the position of the segment points when moving in the smooth
gravitational field of the Galaxy. At the angular velocity of rotation of the spiral
pattern $\Omega_{\text{p}} = 25.2 \pm 0.5$\,km\,s$^{-1}$\,kpc$^{-1}$ (Dambis et al. 2015)
the observed displacement between segments on young and old objects  can be explained by
the amplitude of spiral perturbations of the radial velocity of  $u = 10
\pm1.5$\,km\,s$^{-1}$. For the constructed double system of spiral segments, it is demonstrated that the
assumption of constancy of the pitch angles within each segment and 
the assumption that the pole of the spiral pattern is in the direction 
of the nominal center of the Galaxy do not contradict the data within the range of uncertainty. 
}


\clearpage

\section{Introduction}           
\label{sect:intro}
\noindent
The spiral structure is a significant feature of many disk galaxies. The spiral arms,
that is, the branches of the spiral pattern, are visible against the background of the
galactic disk as narrow elongated areas of increased brightness, delineated by regions of
intense star formation. The presence of spiral arms in the Milky Way has been known since
the mid-20th century (e.g., van de Hulst et al. 1954), but the question of the morphology
of the spiral structure is still not fully clarified (see, e.g., a brief review in
Nikiforov \& Veselova 2018a).

The most common model of a spiral arm is the logarithmic spiral, with~a single value
of the pitch angle~$i$ for all the arms or with a different value for each arm. However,
the number of arms in different models of the Galaxy may be different, and the specified
number of arms largely determines the value of $i$. For example, Francis \& Anderson
(2012) adopted a two-arm logarithmic model of a spiral structure with a constant pitch
angle $i=-5\fdg56\pm0\fdg06$, which was the same for both arms traced by giant molecular
clouds, H\,II regions, 2MASS sources and H\,I distribution for Galactocentric distances up
to $12$--$15$ kpc. On the other hand, Vall\'ee (2008), based on the analysis and
comparison of a number of works, suggested a four-arm logarithmic model with a pitch angle
$i=-12\fdg8$ as the most suitable model of a spiral pattern, consistent with the
tangential directions to the arms determined by~observations of H\,II, $^{12}$CO,
$^{13}$CO,  and~$^{26}$Al. According to a survey in Vall\'ee (2015), published pitch
angle values, depending on the accepted number of arms, cover from $-6\dg$ to $-28\dg$ for
the global Galaxy.

In many studies, the spiral pattern is assumed to be symmetrical: the arms pass into
each other when turning around the Galactic center at an angle of $180^\circ$ in the case
of a two-arm pattern or $90^\circ$ in the case of a four-arm pattern. This assumption,
however, leads to an additional condition: the pitch angles of the arms in this case must
be equal. Efremov (2011) based on the analysis of data on the distribution of neutral,
molecular and ionized hydrogen suggested the presence of a symmetrical logarithmic
four-arm spiral pattern in the inner part of the Galaxy, but noted that the symmetry may
be broken in the outer regions. 

In recent years, the assumption of equal pitch angles for all arms has been gradually
abandoned. For example, in Bobylev \& Bajkova (2014) the value of the pitch angle for four
segments of the arms is estimated based on data on the spatial distribution of masers and
very young open clusters in the Outer arm. The resulting pitch angles of individual segments are
compatible with each other and close to $i = -13^\circ \pm 1^\circ$.

However, the results of a number of studies suggest the possibility of a
significantly more complex morphology. L\'epine et al. (2001) considered the spiral
pattern as a superposition of two- and four-arm models with different pitch angles
($6^\circ$ for two-arm and $12^\circ$ for four-arm); thus, the total number of spiral arms
is six. The proposed model better satisfies the data on Cepheid kinematics than the
two-arm and four-arm models separately and can explain the possible bifurcation of arms,
an example of which can be observed in the galaxy M~101. Englmaier et al. (2011) according
to the data on the distribution of neutral hydrogen suggest the following model of a
spiral pattern: in the inner part of the Galaxy, two significant arms start from the ends
of the bar, then, at a Galactocentric distance of about $7.8~$kpc, the two-arm pattern
splits into four arms, continuing to a distance of about $20~$kpc. The bifurcation points
were not $180^\circ$ apart; one of the~branching points is presumably located near the
Sun, which may affect the velocity distribution in the solar neighborhood.

The problem of determining the parameters of the Galactic spiral structure can also be
associated with the problem of determining the distance to the Galactic center ($R_0$)
under the assumption that the Galactic center is the pole of the spiral arms. Fixing the
value of $R_0$ is a standard assumption in studies of the spiral structure (e.g., in the
mentioned works $R_0=7.5$--$8.0$~kpc). Due to complex correlations with other parameters
of spiral arms (see Nikiforov \& Veselova 2018a), this creates an additional source of
systematic errors, different for different subsystems due to mismatch of methods for
determination of heliocentric distances and mismatches (at least for calibrations) of
photometric distance scales. Freeing~$R_0$ when modeling the spiral structure largely
removes these problems. \linebreak On the other hand, this gives another independent method for
determining $R_0$. The question of a reliable value of $R_0$ cannot be considered closed
with confidence, since modern estimates still differ significantly from each other.
For example, Braga et al. (2018) based on data on Cepheids in the bulge obtained an estimate
of $R_0=8.46 \pm 0.03\,(\text{stat.}) \pm 0.11 \,(\text{sys.})$~kpc; according to data on S0-2
star orbiting the supermassive black hole at the Galactic center GRAVITY Collaboration et
al. (2019) presented an estimate of \linebreak $R_0=8.178 \pm 0.013\,(\text{stat.}) \pm
0.022\,(\text{sys.})$~kpc, and Do et al. (2019) gave a value of $7.946 \pm
0.050\,({\text{stat}}) \pm 0.032\,({\text{sys}})$~kpc.

In our previous works (Nikiforov  \& Veselova 2015, 2018a) we proposed a new
approach to determining the geometric parameters of the Milky Way  spiral arm segments
from the spatial distribution of tracing objects: in an effort to minimize
assumptions, we do not assume or determine the number of spiral arms, but rather estimate
the parameters of \emph{individual} detected segments of arms, considering the geometric
pole to be the same for all segments and obtaining the parameter $R_0$ together with the
geometric parameters, such as the pitch angles and the positional parameters of the
segments. The method was tested on maser source data, and we obtained a solar Galactocentric
distance estimate of $R_0=8.8\pm0.5$\,kpc. Our numerical experiments (Nikiforov  \&
Veselova 2018b) have confirmed the effectiveness of the algorithm developed by us for a
wide range of possible parameter values and made it meaningful to develop a more complex
method that will take into account the influence of distance uncertainties and the natural
scatter of objects across the segments.

Using the maximum likelihood method we developed an algorithm for the spatial modeling of
spiral arm segments taking into account the natural dispersion across the segment and the
uncertainty of the distance modulus, $d$. The algorithm does not require the initial
strict assignment of the object to a specific segment. In Veselova \& Nikiforov (2018) the
proposed method was applied to data on the spatial distribution of classical Cepheids from
the catalog in Mel'nik et al. (2015). The initial assignment of objects to segments was
carried out by analyzing the distribution of $X_{\text{s}}$ coordinates, where $X_{\text{s}}$
is the abscissa of the intersection point of the logarithmic spiral 
corresponding to the object's phase by the direction to the Galactic center.

In this paper, based on the developed algorithm (Sect.~\ref{sect:model}), we find out
whether the age heterogeneity of classical Cepheids is a significant factor in the spatial
modeling of spiral segments traced by these objects, and then investigate the detected age
effects. In~Sect.~\ref{sect:ceph}, we determine the parameters of the spiral structure for
old and young Cepheids separately and together, allowing for a distinction between two
systems of segments traced by young and old objects. Then we perform a dynamic modeling to
interpret the detected displacement between the Perseus and Sagittarius-2 segments in the
two systems (Sect.~\ref{sect:modeling}). In Sect.~\ref{sect:vary}, we show how a fixed
value of $R_0$ affects the estimates of the pitch angles, and test the possibility of
rejecting other standard assumptions when studying the spiral structure---the constancy of
the pitch angles and the coincidence of the arms pole with the Galactic center.


\section{The method of modeling the spiral arm segments}
\label{sect:model}

\subsection{Likelihood function for a set of segments}
\noindent
We investigate the distribution of
objects in the projection on the Galactic plane in the Cartesian coordinate system. The
model of the center line of the segment depends on a number of parameters
($\vec{p}$) 
and represents
the dependence of the Galactoaxial distance on the Galactocentric longitude $\lambda$ measured from the sunward direction clockwise:
\begin{equation}
 R_{\text{mod}} = R_{\text{mod}}(\lambda,\vec{p}).
 \end{equation}
 
\begin{figure}
   \centering
  \includegraphics[width=8cm, angle=0]{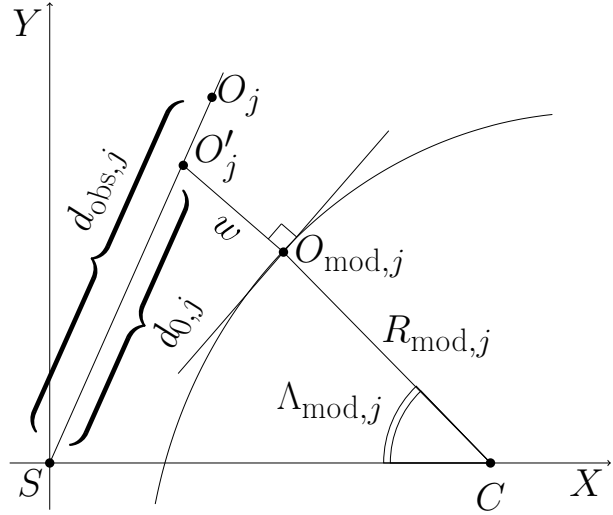}
   \caption{\rm\!Location of an object relative to the center line of a spiral segment
   (curved line).
   The Sun is placed at the origin of the Cartesian coordinate system, $XY$ is the Galactic
   plane and $C$ represents the Galactic center. $O_j$ is the observed position of the $j$-th
   object, $O'_j$ signifies the reduced position of the same object,
   $O_{\text{mod},j}$ is the point on a center line for which the distance $w$ from
   reduced position is minimal (the ``model position'' of the object) and
   ${\Lambda_{\text{mod},j}}$ is the Galactocentric longitude
   of the point $O_{\text{mod},j}$.} 
   \label{scheme}
   \end{figure}

We assume that the displacement $w$ of the object across the segment (for which the shape of the center line is given by
the model) and the error of the distance modulus obey the normal distribution, so the
likelihood function is the product of the corresponding distribution functions, taking
into account the uniform dispersion $\sigma_{\text{w}}$ across the arm and the uncertainty
of the distance modulus $\sigma_d$. For one segment, the likelihood function is given as
\begin{equation}
 L = \prod\limits_{j=1}^N \frac{1}{\sqrt{2\pi}\sigma_{w}}\exp\left[-\frac{w^2(d_{0,j},\vec{p})}{2\sigma^2_{w}}\right]
   \frac{1}{\sqrt{2\pi}\sigma_{d}}\exp\left[-\frac{(d_{\text{obs},j} - d_{0,j})^2}{2\sigma^2_{d}} \right],
\end{equation}
but for convenience  we use the negative log likelihood function
$\mathfrak{L}$ represented as
\begin{equation}
 \mathfrak{L} = -  \ln L   = N\ln(2\pi) + N\ln\sigma_{d} +   N\ln\sigma_{w} 
   + \frac{1}{2}\sum\limits^N_{j=1}\underbrace{\min_{d_{0,j}}\left[ \frac{w^2(d_{0,j},\vec{p})}{\sigma_{w}^2} +
  \frac{(d_{\text{obs},j} - d_{0,j})^2}{\sigma^2_{d}}\right]}_{\Large {S^2_j}},
\end{equation}
where $N$ is the number of objects, $d_{\text{obs},j}$ 
represents the observed (measured)
distance modulus of the $j$-th object and $d_{0,j}$ is its reduced distance modulus (see
Fig.~\ref{scheme}).
  The expression in square brackets can be considered as the
square of weighted distance from the reduced position of the 
$j$-th object to the center line of the segment at the accepted value of
$d_{0,j}$. The calculation  of the segment parameters is performed as follows: first, we
find the minimum value~$S^2_j$ of the square of weighted distance
 by varying the reduced
   distance modulus  $d_{0,j}$ 
 for each $j$-th object at a given
 set of parameters $\vec{p}$, i.e. we determine the reduced position  $O'_j$ of an observed position $O_j$ of the $j$-th  object  by
 shifting it on the line of sight so that for  $O'_j$ 
 the weighted distance is
 minimal, $O_{\text{mod}, j}$ 
 is the point on a center line for which the distance
 $w$ from reduced position is minimal, in fact, this is a projection of the reduced
 position $O'_j$  on the model spiral.  Next we can determine the overall minimum
 $\mathfrak{L}_0$ for the segment while minimizing its parameters $\vec{p}$. The shape of
 the center line of the segment determines the value of the weighted distance.

When considering a set of segments, the likelihood function takes the form of a
product of functions for individual segments.  We assume  $\sigma_d$ to be the same for
the entire sample of objects, but $\sigma_w$ has its own value ($\sigma_{{w},a}$) for each
segment. For a set of $N_{\text{segm}}$ segments the negative log likelihood
function is represented as
  \begin{equation}\label{fun}
 \hskip-0.3cm 
 \mathfrak{L}  = N\ln(2\pi) + N\ln\sigma_{d} +
    \sum\limits_{a=1}^{N_{\text{segm}}}\!\!N_a\ln\sigma_{{w},a} +
    \frac{1}{2} \sum\limits_{a=1}^{N_{\text{segm}}}\sum\limits^{N_a}_{j=1}{\min_{d_{0,j}}
	    \left[ \frac{{w}^2(d_{0,j}\,, \vec{p_a})}{\sigma_{{w},a}^2} +
    \frac{(d_{\text{obs},j} - d_{0,j})^2}{\sigma^2_{d}}\right]}.
   \end{equation}
Here $N$ is the total number of objects, $N_a$ is the number of objects tracing the $a$-th segment and $\vec{p_a}$  is the set of parameters of the $a$-th segment
\newpage 

\subsection{Likelihood function in case of a logarithmic model}
\noindent
The assumption about the model of the segment center line is applied when we calculate the weighted distances. We have chosen the logarithmic spiral as the most popular model. We 
assume the spiral to be in the Galactic plane and 
the direction from the Sun to the spiral pole (to the 
Galactic center) to be known. The  spiral 
arm is then  represented by a segment of the 
logarithmic spiral 
\begin{equation}
 R_{\text{mod}}(\lambda; R_0, k, X_{\text{s}}) = |R_0-X_{\text{s}}|\mathrm{e}^{k\lambda}.
\end{equation}
Here, $\lambda \in  ( - \infty, + \infty )$ is the Galactocentric longitude (it is measured from the sunward direction clockwise
when viewed from the North Galactic Pole); $k \equiv \tan{i}$, where $i$ is
the pitch angle (it is negative for a trailing segment); $X_{\text{s}}$ is the abscissa of the point, at which  the segment intersects the direction to the Galactic center. In the sunward
direction $\lambda = 0 \pm 2\pi n, n \in \mathbb{Z}$.

In general, we assume a value of the solar Galactocentric distance $R_0$ to be common
for all segments, but the pitch angles, $X_{\text{s}}$ values and dispersions
$\sigma_{w}$  may differ for different segments. For a logarithmic model the
likelihood function  $\mathfrak{L}$ for a set of segments takes the form
  \begin{equation}
  \begin{split}
  \mathfrak{L}  = N\ln(2\pi) + N\ln\sigma_{d} &+ \sum\limits_{a=1}^{N_{\text{segm}}}N_a\ln\sigma_{{w},a} + \\
 & +  \frac{1}{2}\sum\limits_{a=1}^{N_{\text{segm}}}\sum\limits^{N_a}_{j=1}{\min_{d_{0,j}}
	\left[ \frac{{w}^2(d_{0,j}\,;R_0, i_a, X_{\text{s},a})}{\sigma_{{w},a}^2} +
	\frac{(d_{\text{obs},j} - d_{0,j})^2}{\sigma^2_{d}}\right]}.
	\end{split}
   \end{equation}

To determine the distance $w$ from the reduced position $O_j'$ of the $j$-th object
with the coordinates $(X_{0,j}, Y_{0,j})$ to the spiral, the roots of the following
transcendental equation on the longitude $\Lambda_{\text{mod},\,j}$ of the point on the
spiral with the minimum distance to $O_j'$ are calculated:
\begin{equation}\label{eqdist}
  \begin{split}
(X_{0,j} - R_0 )\big(\sin{\Lambda_{\text{mod},\,j}} - k_a\cos{\Lambda_{\text{mod},\,j}} \big)  - k_a&|R_0-X_{\text{s},a}|\mathrm{e}^{k_a{\Lambda_{\text{mod},\,j}}} + \\ 
& + Y_{0,j}\left(k_a\sin{\Lambda_{\text{mod},\,j}} +  \cos{\Lambda_{\text{mod},\,j}} \right)  = 0,
\end{split}
\end{equation}
where $k_a\equiv \tan i_{0,a}$. 
Optimal parameters of a set of segments are determined by minimizing the
function~$\mathfrak{L}$. In this way, we jointly obtain the values of $R_0$, pitch angles
$i_a$, parameters $X_{\text{s},a}$ and dispersions $\sigma_{w,a}$ for $a=
1,2,\,\ldots,\,N_{\text{segm}}$. Generally  the number of optimized values is $M =
3N_{\text{segm}}+1$.

In the case of far or sparsely populated segments the deviation of objects from the
center line of the segment can be completely explained by only the presence of uncertainty
in the distance moduli. For these segments we set $\sigma_w = 0$ and the negative
likelihood function takes the form
\begin{equation}
 \mathfrak{L} =  \frac{N}{2}\ln(2\pi) + N\ln\sigma_{d}  
   + \frac{1}{2}\sum\limits^N_{j=1}\underbrace{\min_{d_{0,j}}
  \frac{(d_{\text{obs},j} - d_{0,j})^2}{\sigma^2_{d}}}_{\Large {S^2_j}}.
\end{equation}

\subsection{Estimation of confidence intervals}
\noindent
The boundaries of the confidence interval of the parameter $p_j$ for the confidence
level $1\sigma$ can be determined from the equation
\begin{equation}
 \mathfrak{L}_{\text{m}}(p_j) = \mathfrak{L}_0 + \frac12\,, \qquad
 \mathfrak{L}_0 = \min \mathfrak{L}, \qquad
 \mathfrak{L}_{\text{m}}(p_j)= \min_{p_j=\text{const}} \mathfrak{L}.
\end{equation}
Here $\mathfrak{L}_0$ 
is the minimum of the likelihood function, and $\mathfrak{L}_{\text{m}}$ is the profile of
the log  likelihood function for the parameter $p_j$ obtained by optimizing all
parameters except $p_j$.

To reduce the computation time in our work, the boundaries of confidence intervals
were determined in the parabolic approximation.  We consider the deviation $\Delta p_j>0$
of the parameter $p_j$ from the optimal value $p_j^0$, then determine the values of
$\mathfrak{L}_{\text{m}}$ at fixed $p_j^-= p_j^0 - \Delta p_j$ and $p_j^+= p_j^0 + \Delta
p_j$, and then find the differences $\sigma^+$ and $\sigma^-$ between the optimal value
$p_j^0$ and the boundaries of the confidence interval
\begin{equation}
 \sigma^{-} =\frac{\Delta^2 p_j}{2\big(\mathfrak{L}_{\text{m}}(p_j^-) - \mathfrak{L}_0\big)}, \quad  \sigma^{+} =\frac{\Delta^2 p_j}{2\big(\mathfrak{L}_{\text{m}}(p_j^+) - \mathfrak{L}_0\big)}.
\end{equation}

\subsection{Assignment of objects to spiral segments}
\noindent
There may be a question about the correct assignment of objects to spiral segments. We
have proposed and implemented (Veselova
\& Nikiforov 2018) the following algorithm. The initial division of objects into segments is carried out in accordance with the
minima of the distribution function of the values of $X_{\text{s}}$ which are computed
for every object under the assumption of $R_0$ and $i_0$ values on a grid. Next
 we identify the basic sample of objects whose assignment to segments
does not depend on assumptions on~$R_0$. Then we optimize the parameters of segments
traced by the basic sample by minimizing $\mathfrak{L}$, 
and other objects are assigned to the segments in accordance with the minimum weighted
distance: for each object we compute the $S_j^2$ values relative to each segment and
assign the object to the segment with minimal $S_j^2$. Next, the iterative reassignment of
objects to specific segments and the optimization of parameters for the final assignment of
objects into segments are carried out. 

\begin{figure}[!tp]
   \centering
  \includegraphics[width=7.8cm, angle=0]{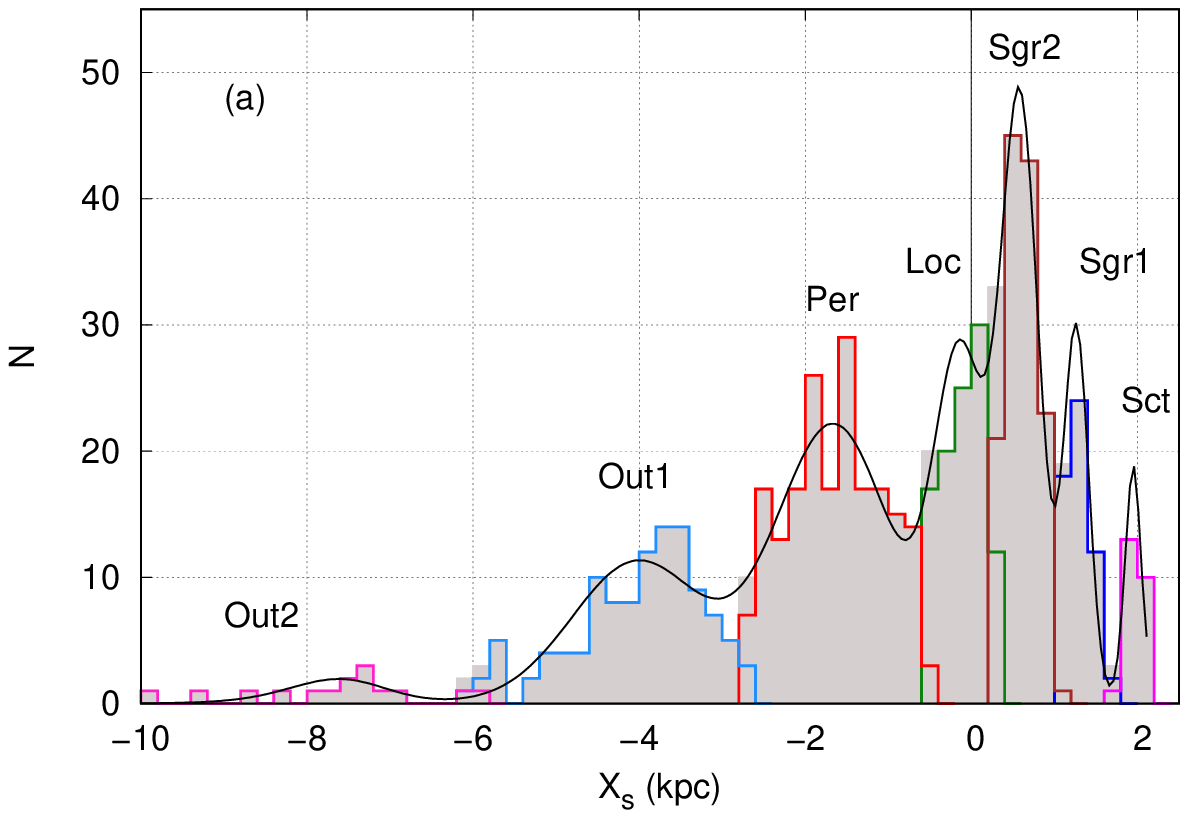}
  \includegraphics[width=7.8cm, angle=0]{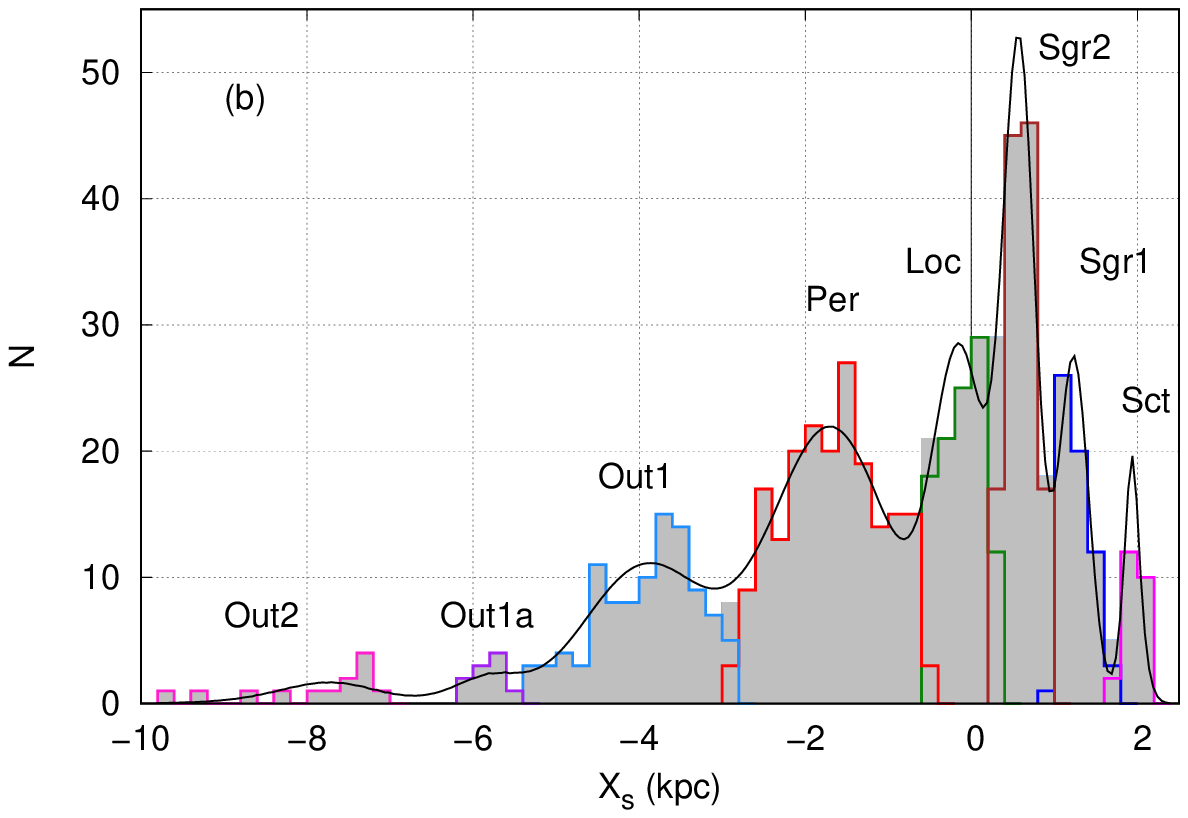}
  \includegraphics[width=7.8cm, angle=0]{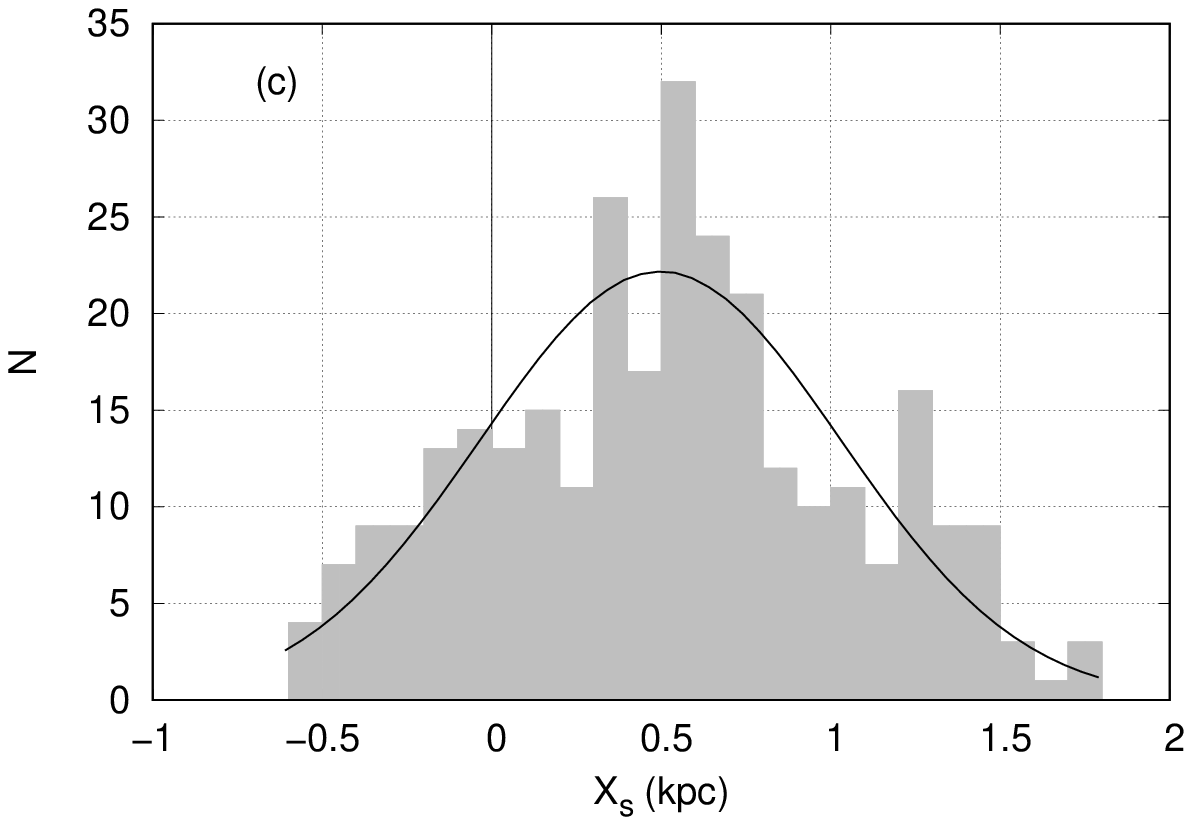}
  \includegraphics[width=7.8cm, angle=0]{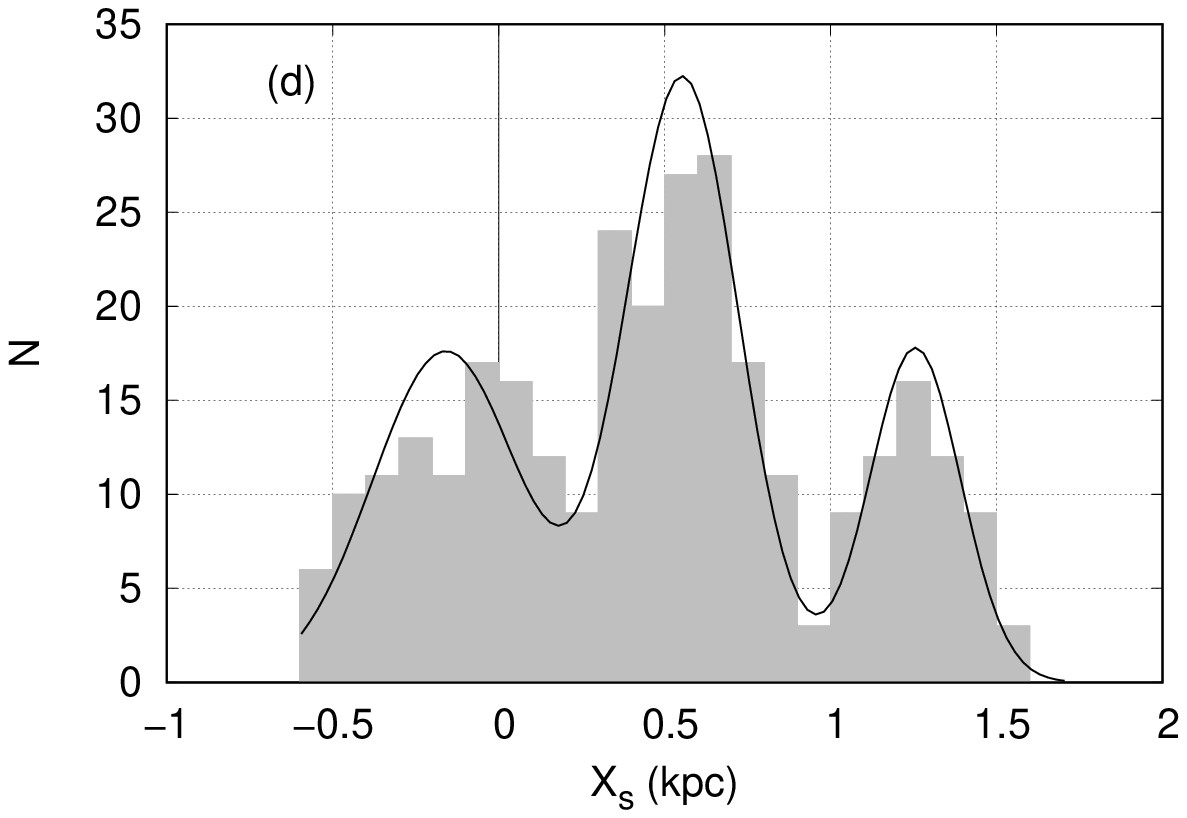}
   \caption{\rm\!Distribution of $X_{\text{s}}$ values in the cases of seven (a) and eight (b) segments
in the model, taken from Veselova \& Nikiforov (2018); the $X_{\text{s}}$ distributions
in the case of a single Sagittarius arm (c) and when dividing it into
three separate segments (d), constructed only from the Cepheids tracing these details.
The
solid line displays the model function of the distribution of values $X_{\text{s}}$ calculated with
the uncertainty in distance moduli taken into account.  In each panel, the position of the Sun
corresponds to $X_{\text{s}} = 0.0$ kpc as marked by the solid vertical line.} 
   \label{xs}
   \end{figure}

We chose classical Cepheids as objects that trace the spiral structure. In Veselova
\& Nikiforov (2018) and in current work we utilized data from the catalog in Mel'nik et al.
(2015) (see Sect.~\ref{sect:ceph} for details).  Analysis of the spatial distribution of
565 Cepheids from this catalog allowed Dambis et al. (2015) to identify four segments of
the global spiral structure. In Veselova
\& Nikiforov (2018) we aimed to undertake a more detailed analysis of the nearest spiral
structure and we also considered slightly larger region populated by Cepheids.  According
to the algorithm described above, we investigated the location of minima of the
distribution function of $X_{\text{s}}$ values for different values of $R_0$ and for models
with different numbers of segments. Figure~\ref{xs}(a) and \ref{xs}(b) depicts the $X_{\text{s}}$ 
distribution for models with seven and eight segments.
The analysis of the consistency of the model and observed distribution functions,
conducted applying the Pearson ($\chi^2$) criterion, indicated the preference for a model with eight
segments.

Thus, according to the data on 636 Cepheids, we preferred an eight segment model of nearby
spiral structure. In comparison with the set of segments presented in Dambis et al.
(2015), we divided the single Sagittarius arm into three segments (Local arm,
Sagittarius-1 and Sagittarius-2) and, considering a slightly larger spatial region
populated by Cepheids, we also added a more distant arm (Outer-2) and identified a small
segment (Outer-1a). In Figure~\ref{xs}(c) and~\ref{xs}(d), we compare the model with a single
Sagittarius arm and the model with three distinct segments. According to the Pearson
criterion, the model with a single segment is rejected.  Note that we did not draw
conclusions about the global spiral structure, and only discussed the parameters of
individual segments.

\section{Determination of parameters of segments traced by classical Cepheids}
\label{sect:ceph}
\noindent
In this paper we compare the parameters of spiral segments traced by young and old
classical Cepheids from the catalog in Mel'nik et al. (2015). This catalog contains
data on coordinates, proper motions, radial velocities, pulsation periods, and apparent
stellar magnitudes of 674 Cepheids. Authors did not specify the possible value of the
distance modulus error. In our work we assumed $\sigma_d = 0.14^m$, based on comparisons
with other distance catalogs, as well as a sharp restriction from below on $\sigma_d$
created by a very narrow Scutum arm. This value allowed us to determine the value of
dispersions across the arm in fairly populated segments. We discuss the possible influence
of adopted $\sigma_d$ on parameters of segments later in this section.

According to the distance scale utilized in the catalog, the distance modulus  of the
Large Magellanic Cloud (LMC) is $d_{\text{LMC}} = 18.25\pm0.05$ (Berdnikov et al. 2000). We
rescale the final $R_0$ estimate obtained by us to the modern LMC calibration
$d_{\text{LMC}} = 18.49\pm0.09$ (de Grijs et al. 2014), i.e. we use a correction factor of
1.117 (Sect.~\ref{sect:conclusions}). Other parameters of spiral segments obtained in this
work should also be multiplied by this factor.

\begin{figure}[!t]
   \centering
  \includegraphics[width=9.5cm, angle=0]{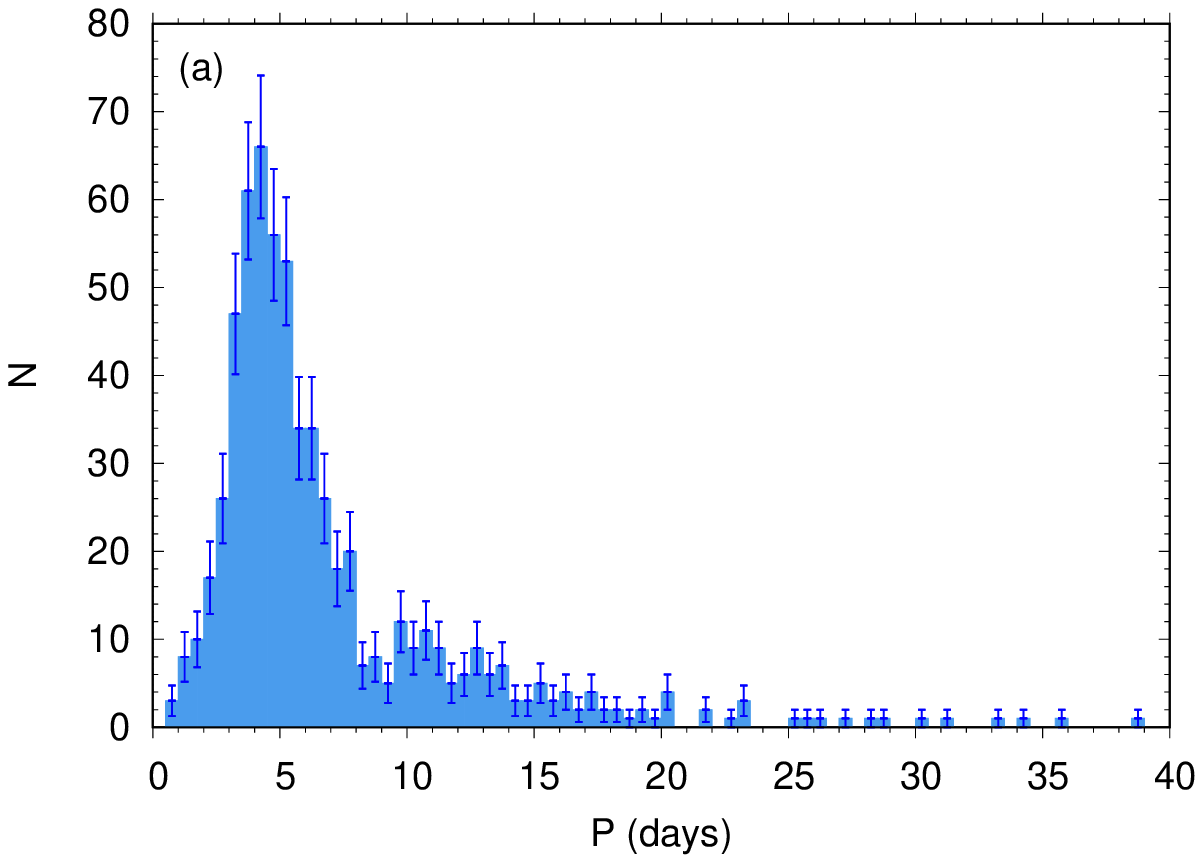}
  \hskip-0.3cm
  \includegraphics[width=6.6cm, angle=0]{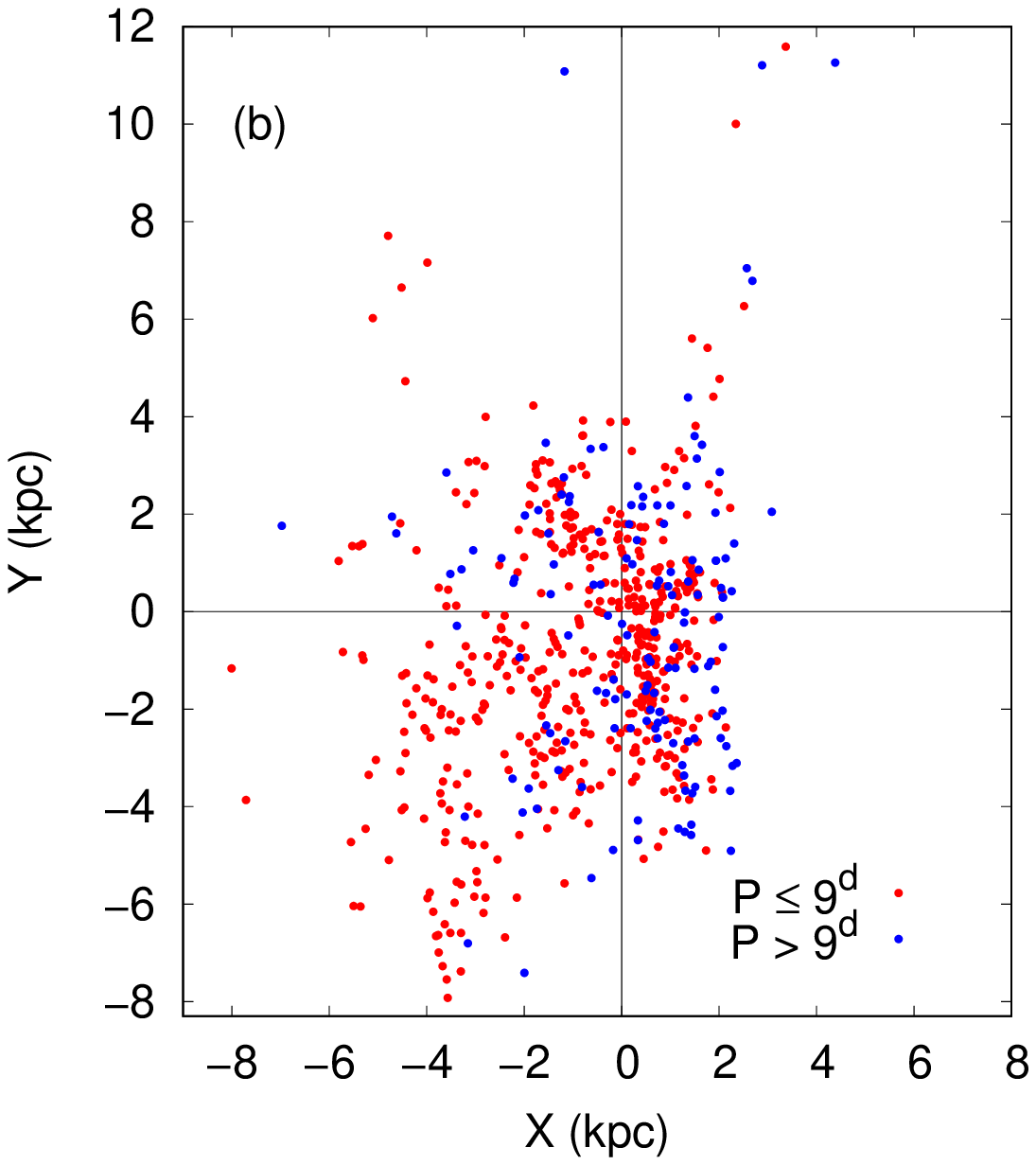}
   \caption{\rm\!Distribution of pulsation periods (a) and spatial distribution of old and
   young Cepheids projected on the Galactic plane (b). Old Cepheids are displayed in red
   and young Cepheids are marked in blue. The Sun is placed at $X = 0.0$ kpc, $Y = 0.0$ kpc.} 
   \label{age}
   \end{figure}

As had been demonstrated by Karimova \& Pavlovskaya (1974), the distribution of classical
Cepheids over pulsation periods has a minimum near the period of $9^\text{d}$ (see
also Fig.~\ref{age}a, based on the data we used). Cepheids with smaller values of the
period have a greater age, then we call the selection of such objects old, and the rest of
the objects are called young. In total, the sample considered consisted of $140$ young and $494$
old objects.

According to different relations for the age dependence on the pulsation period, we
get slightly different age estimates for the  period of $9$ d. According to the relation in
Efremov (2003), the age estimate is ${\sim}7.6\times10^7$ yr. If we apply the relation in
Joshi \& Joshi (2014), we obtain an estimate of ${\sim}7.3\times10^7$ yr. 
The median pulsation period for the old sample is nearly $4.6$ d, and if we consider 
several modern relations, we conclude that this value corresponds to the age of $1\times10^8$ yr, 
and the median period for the young sample---$13.5$ d---corresponds to the age of  $5\times10^7$ yr.


The spatial distribution of young and old Cepheids is presented in Fig.~\ref{age}(b). We
can see that young objects are situated mostly inside the solar circle, but outside the
solar circle they are located very sparsely. The nearest to the Galactic center segment
(Scutum arm) consists mostly of young Cepheids, but the Outer segments are mostly represented
by old objects.

In accordance with the previously mentioned algorithm, the samples were divided into
segments. It was revealed that young objects trace only seven separate segments (there is no
segment Outer-1a), and old objects form eight segments. Note that when optimizing the
parameters separately for each of two samples, the estimates of $R_0$ are similar: $R_0 =
7.15 \pm 0.24$\,kpc for young objects and $R_0 = 7.33\pm0.29$\,kpc for old objects.
The similarity of these values provides a basis for joint optimization of the
parameters of two spiral patterns traced by young and old Cepheids. The results of the
simultaneous optimization of the parameters for two samples are presented in
Table~\ref{tabpar}, and the value $R_0$ is equal to $7.23_{-0.18}^{+0.19}$\,kpc. 
Table~\ref{tabdiff} lists the differences between the parameters of segments traced
by old and young Cepheids. Difference $\Delta i_{\text{o}-\text{y}}$ 
is significant for  the Sagittarius-1 and Sagittarius-2 arms, for the Local arm
significance is marginal because $\Delta i_{\text{o}-\text{y}}$ is only greater than $2\sigma$.
Difference  $\Delta X_{\text{s},\text{o}-\text{y}}$ in the $X_{\text{s}}$ values is significant for
the Sagittarius-2 arm, but for the Local and Perseus arms significance is marginal.  

\begin{table}
\caption[]{\rm\!Parameters of spiral segments traced by young and old Cepheids. A zero
width value $\sigma_w$ indicates that the deviation of the object from the center line of the segment
can be explained by the uncertainty in distance moduli\label{tabpar}}
\tabcolsep0.23cm
\small
 \begin{tabular}{l|ccc|ccccccccccccc}
  \hline\noalign{\smallskip}
   & &  Young  & &  & Old &  \\
  Segment & $i$ & $X_{\text{s}}$ (kpc) & $\sigma_w$ (kpc) & $i$ & $X_{\text{s}}$ (kpc) & $\sigma_w$ (kpc) \\[0.1cm]
  \hline\noalign{\smallskip}
Scutum & $-11\fdg7^{+0\fdg9}_{-0\fdg9}$ & $2.01^{+0.04}_{-0.03}$ & $(0.00)$  & $-8\fdg9^{+2\fdg1}_{-2\fdg1}$ & $1.90_{-0.06}^{+0.06}$ & $(0.00)$\\[0.1cm]
Sagittarius-1 & $-13\fdg1^{+1\fdg4}_{-1\fdg4}$ & $1.30_{-0.05}^{+0.05}$ & $0.214_{-0.032}^{+0.033}$ &  $-8\fdg3^{+0\fdg8}_{-0\fdg8}$ & $1.25_{-0.02}^{+0.02}$ & $0.119_{-0.017}^{+0.017}$\\[0.1cm]
Sagittarius-2 & $-10\fdg9_{-0\fdg6}^{+0\fdg6}$ & $0.699_{-0.022}^{+0.023}$ & $0.065_{-0.013}^{+0.013}$ & $-7\fdg4_{-0\fdg7}^{+0\fdg7}$ & $0.600_{-0.020}^{+0.020}$ & $0.144_{-0.011}^{+0.011}$ \\[0.1cm]
Local &  $-9\fdg9_{-1\fdg2}^{+1\fdg2}$ & $-0.111_{-0.048}^{+0.048}$ & $0.239_{-0.033}^{+0.033}$ & $-7\fdg0_{-0\fdg6}^{+0\fdg6}$ & $0.002_{-0.023}^{+0.023}$ & $0.174_{-0.016}^{+0.016}$\\[0.1cm]
Perseus & $-6\fdg2_{-1\fdg6}^{+1\fdg6}$ & $-1.74_{-0.09}^{+0.09}$ & $0.490_{-0.065}^{+0.066}$ & $-7\fdg0_{-1\fdg2}^{+1\fdg2}$ & $-1.54_{-0.05}^{+0.05}$ & $0.630_{-0.035}^{+0.036}$ \\[0.1cm]
Outer-1 & $-5\fdg2_{-2\fdg8}^{+2\fdg8}$ & $-3.94_{-0.17}^{+0.17}$ & $0.437_{-0.147}^{+0.146}$  & $-9\fdg7_{-1\fdg1}^{+1\fdg1}$ & $-3.85_{-0.07}^{+0.07}$ & $0.368_{-0.055}^{+0.055}$ \\[0.1cm]
Outer-1a & & & & $-8\fdg1_{-1\fdg7}^{+1\fdg7}$ & $-5.81_{-0.13}^{+0.13}$ & $(0.00)$ \\[0.1cm]
Outer-2 & $-8\fdg3_{-2\fdg6}^{+2\fdg6}$ & $-7.54_{-0.55}^{+0.54}$ & $(0.00)$ & $-9\fdg2_{-2\fdg0}^{+2\fdg0}$ & $-7.70_{-0.27}^{+0.27}$& $(0.00)$ \\[0.1cm]
  \noalign{\smallskip}\hline
\end{tabular}
\end{table}

\begin{table}
\caption[]{\rm\!Differences between the parameters of segments traced by
old and young Cepheids.  $N_{\text{o}}$ and $N_{\text{y}}$ denote the numbers of objects
that trace the old and young segments, respectively. $\Delta i_{\text{o}-\text{y}}$ expresses the
difference between the pitch angles and $\Delta X_{\text{s},\text{o}-\text{y}}$ signifies the
difference between the $X_{\text{s}}$ values of the old and young
segments\label{tabdiff}} 
 \small
 \centering
 \tabcolsep0.43cm
 \begin{tabular}{lcccccccccccc}
  \hline\noalign{\smallskip}
  Segment & $N_{\text{o}}$ &  $N_{\text{y}}$ & $\Delta i_{\text{o}-\text{y}}$ & $\Delta X_{\text{s},\text{o}-\text{y}}$ (kpc) \\
  \hline\noalign{\smallskip}
Scutum & $6$ & $17$& $2\fdg8\pm2\fdg3$& $-0.11\pm0.07$\\[0.1cm]
Sagittarius-1 & $44$ & $27$ & $4\fdg8\pm1\fdg6$ & $-0.05\pm0.06$ \\[0.1cm]
Sagittarius-2 & $94$ & $21$ & $3\fdg5\pm0\fdg9$ & $-0.099\pm0.030$ \\[0.1cm]
Local & $66$ & $28$ & $2\fdg9\pm1\fdg3$ & $+0.113\pm0.053$ \\[0.1cm]
Perseus & $178$ & $33$ & $-0\fdg8\pm2\fdg0$ & $+0.20\pm0.10$\\[0.1cm]
Outer-1 & $89$ & $10$ & $-4\fdg5\pm3\fdg0$ & $+0.09\pm0.18$\\[0.1cm]
Outer-2 & $7$ & $4$ & $-0\fdg9\pm3\fdg3$ & $-0.16\pm0.60$ \\[0.1cm]
  \noalign{\smallskip}\hline
\end{tabular}
\end{table}

\begin{figure}
   \centering
  \includegraphics[width=16cm, angle=0]{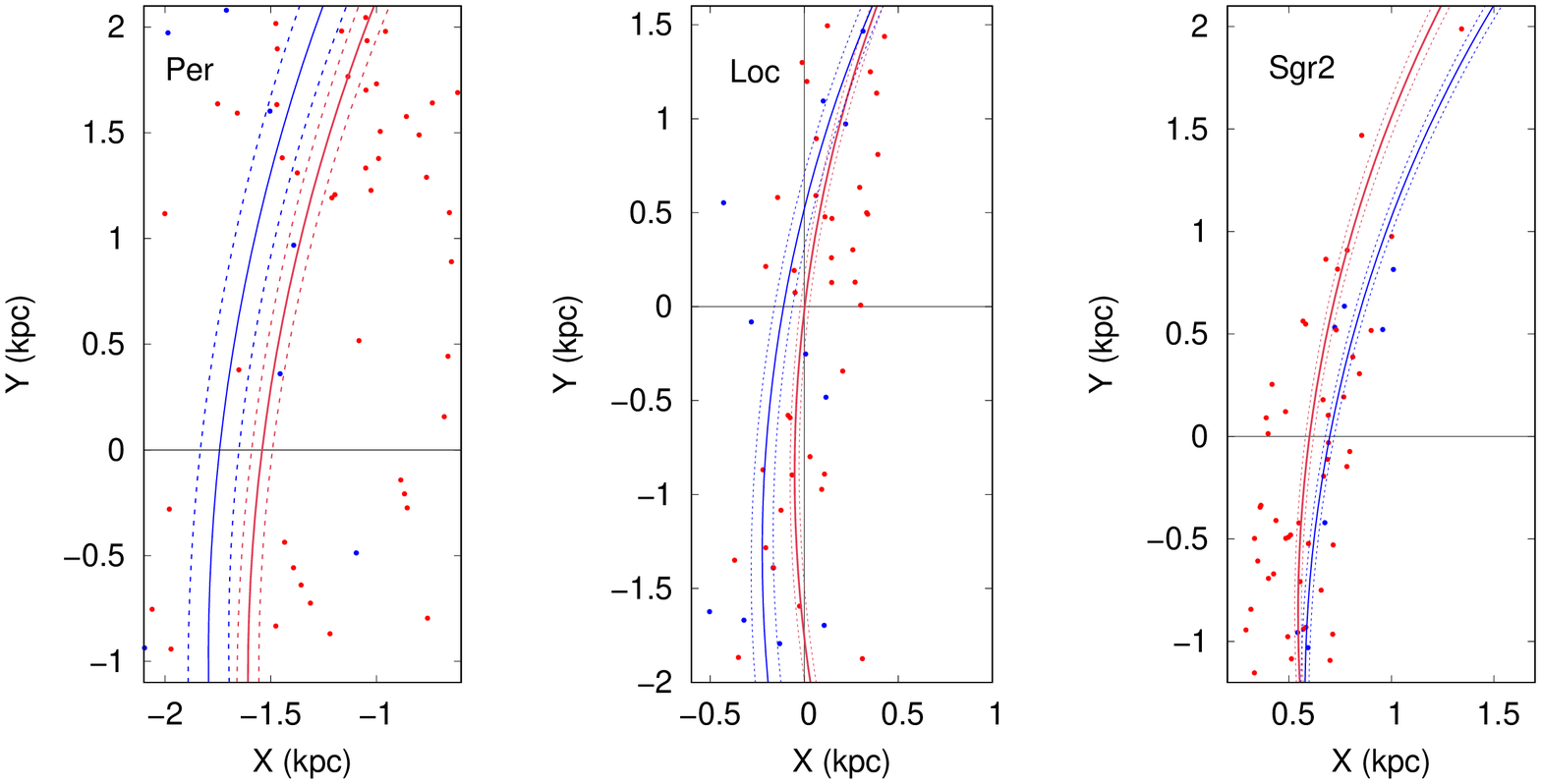}
   \caption{\rm\!1$\sigma$ confidence regions for the Perseus, Local and Sagittarius-2 arms. 
   Solid lines signify the center lines of segments, and dashed lines represent the boundaries
   of confidence regions. Old objects are displayed in red colour and young objects are shown
   in blue colour.} 
   \label{conf}
   \end{figure}

In Figure~\ref{conf} we compare the 1$\sigma$ confidence regions for old and young
segments. We consider only those segments that are quite large and have significant
differences in parameters of center lines. One can see that for all three cases confidence
regions do not intersect near the $X$-axis, so we can say that the young and old segments
differ significantly. 

\begin{figure}
   \centering
  \includegraphics[width=13cm, angle=0]{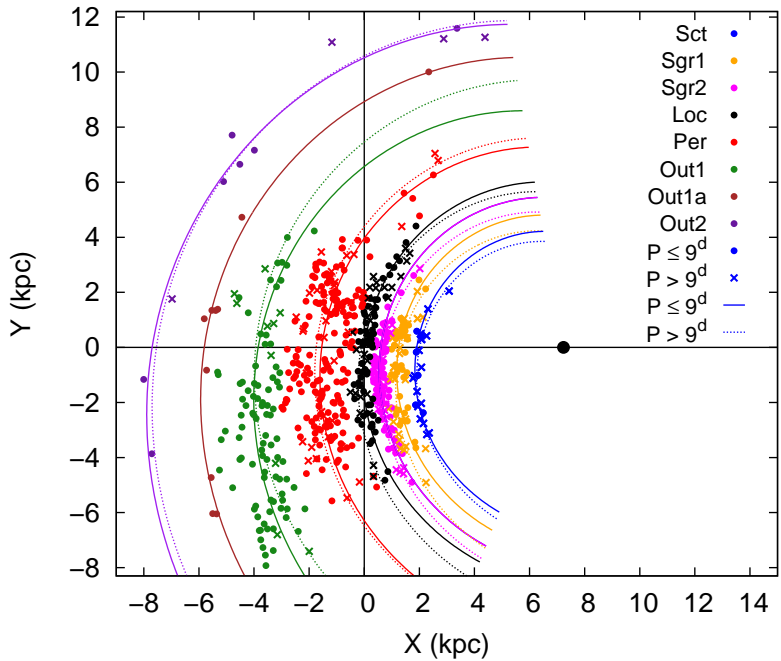}
   \caption{\rm\!Distribution of Cepheids and spiral segments projected on the Galactic
   plane ($R_0 = 7.23$\,kpc, see text). The young sample is represented by
   crosses, the old sample is marked by circles. Segments traced by young objects are
   represented by dotted lines,	while those formed by the old sample are signified by solid
   lines. The Sun is placed at $X = 0.0$ kpc, $Y = 0.0$
kpc. The Galactic Center is located at $X = 7.23$ kpc, $Y = 0.0$ kpc.}
   \label{segm} 
   \end{figure}

The center lines of  all traced  segments in the projection on the $XY$
plane are presented in Figure~\ref{segm}. Dashed lines correspond to segments traced by young
objects, solid lines---for the old ones. One can see that in some cases the center lines
of the young and old segments intersect in the region inhabited by Cepheids.

\begin{figure}
   \centering
  \includegraphics[width=12cm, angle=0]{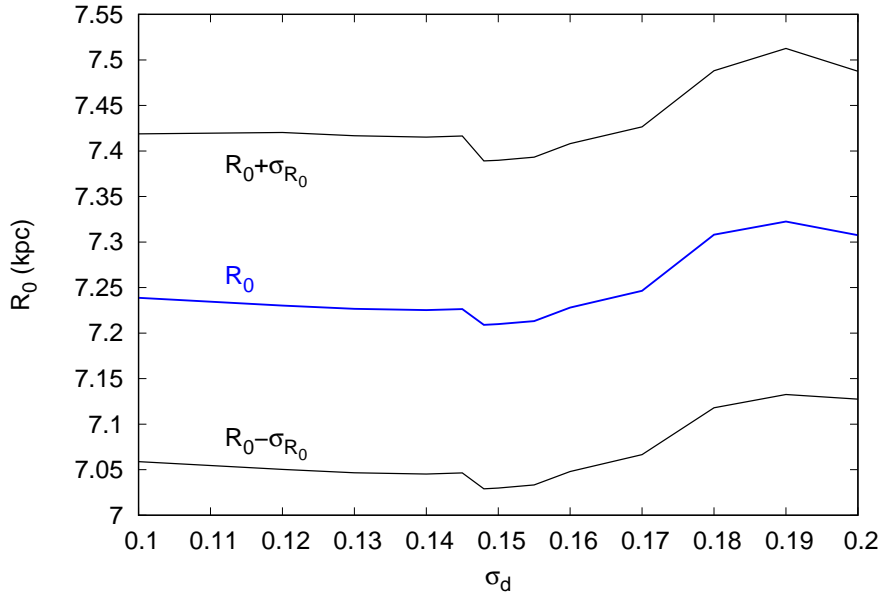}
   \caption{\rm\!Dependence of $R_0$ estimate on the assumed value of uncertainty of
   distance moduli. Blue line indicates the estimate obtained by optimizing parameters jointly
   for all segments, while black lines highlight the boundaries of the confidence interval of the $R_0$
   estimate.} 
   \label{infl}
   \end{figure}

Figure~\ref{infl} shows the dependence of the $R_0$ estimate on the assumed value of
uncertainty $\sigma_d$ of the distance moduli. One can see that the variation in the $R_0$
value does not exceed the statistical uncertainty of $R_0$~estimates. The $X_{\text{s}}$ parameters
and pitch angles for all segments also excibit no significant dependence on $\sigma_d$, 
therefore, we may conclude that the parameters of the center lines of segments are
almost independent of the specific assumption about $\sigma_d$.

\section{Modeling the motion in a smooth Galactic potential}
\label{sect:modeling}
\begin{figure}
   \centering
  \includegraphics[width=12cm, angle=0]{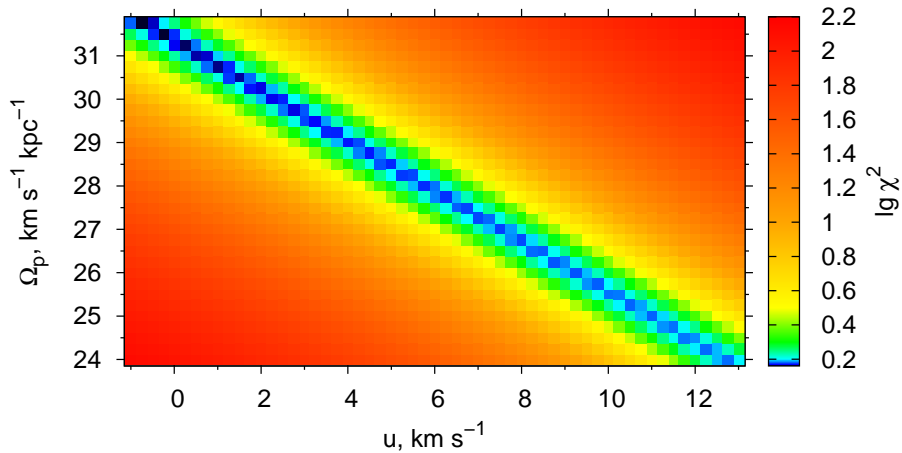}
  \vskip-1.2cm
   \caption{\rm\!Dependence of the decimal logarithm of the value $\chi^2$ on the values of the
angular velocity of the spiral pattern $\Omega_{\text{p}}$ and the component $u$ of the
perturbation velocity of the spiral pattern for the Sagittarius-2 segment.} 
   \label{Om_u_chi}
   \end{figure}

\noindent
In order to explain the bias between the spiral segments traced by young and old Cepheids,
we modeled the motion of objects which were formed in a spiral arm segment.
We chose the smooth potential of the Galaxy considered by Casetti-Dinescu et al. (2013).
In that work, the disk was represented by the Miyamoto--Nagai model, the halo potential had
a logarithmic dependence on a Galactocentric distance and the bar of the Galaxy was
reproduced as an triaxial ellipsoid (the Ferrers potential) with a density profile varying
by the power law. Casetti-Dinescu et al. (2013) implemented the Hernquist potential for
reproducing the bulge component of the Galaxy. In Nikiforov \& Veselova (2020, in prep.)
we have shown that this kind of model is not convenient for our investigation, so we
considered three other kinds of bulge potential models. We chose the Plummer sphere
(Kondrat'ev \& Orlov 2008), the Miyamoto-Nagai disk (Ninkovi\'c 1992) and an isochrone
potential. We tried to estimate the speed of a spiral pattern $\Omega_{\text{p}}$ and the
radial component $u$ of velocity  given to a object by a spiral wave. For a square grid of the parameters in question, one calculates a $\chi$-squared function
for the biases in longitude $\lambda_{\text{int}}$ of the intersection of the young segment and the old
segment and in difference between the $X_{\text{s}}$ values for segments traced by young and old
objects:
\begin{equation}
\chi^2(u, \Omega_{\text{p}}) = \frac{(\Delta X_{\text{s},\text{obs}}-\Delta X_{\text{s},\text{mod}})^2}{\sigma^2(\Delta X_{\text{s}})} +\frac{(\lambda_{\text{int},\text{obs}}-\lambda_{\text{int},\text{mod}})^2}{\sigma^2(\lambda_{\text{int}})}\,.
\end{equation}
We ascertained that chosen kind of a bulge potential does not have a strong influence on
results of modeling because the segments considered are not situated close to the
central part of the Galaxy. We performed numerical simulations for two large
segments. For both Sagittarius-2 and Perseus segments $\Omega_{\text{p}}$ and $u$ are
strongly correlated and could not be estimated simultaneously. The dependence of $\chi^2$
on $\Omega_{\text{p}}$ and $u$ for the Sagittarius-2 segment is depicted in
Figure~\ref{Om_u_chi}. One can see that if we take $u$ equal to~0, we obtain too high
value of $\Omega_{\text{p}}$ (${\sim} 31$\,km\,s$^{-1}$\,kpc$^{-1}$). If we take an
estimate of $\Omega_{\text{p}} = 25.2\pm0.5\,$km s$^{-1}$ obtained by Dambis et al.
(2015), we get the value of $u$ equal to $10\pm1.5$\,km\,s$^{-1}$.

\section{Variability of the pitch angles, influence on the pitch angles of~the~Sun-Galactic center distance, and the displacement of the~pole of~the~spiral pattern}
\label{sect:vary}
\noindent
Savchenko and Reshetnikov (2013) concluded that approximately 2/3 of spiral galaxies manifest
variations in the pitch angle of more than $20\%$, so we investigated the question of  
variability in the pitch angle throughout the spiral segment. We considered the pitch
angle linearly varying with the Galactocentric longitude $\lambda$: 
$i(\lambda)=i_0+i_1\cdot\lambda$.
The only part of the likelihood function
that is to be changed concerns the distance $w$ from a point to the center line of a
segment. The equation for a longitude $\Lambda_{{\text{mod},\,j}}$ 
of a point
representing the base of a perpendicular drawn onto a center line of the $a$-th segment takes the
form
\begin{equation}
\begin{split}
&\left(X_{0,j} - R_0 \right)\left(\sin{\Lambda_{\text{mod},\,j}} - \left[k_a + \frac{i_{1,a}\cdot\Lambda_{{\text{mod},\,j}}}{\cos^2i_a} \right] \cos{\Lambda_{\text{mod},\,j}} \right)  
 +   Y_{0,j}\Big(\cos{\Lambda_{\text{mod},\,j}} +   \label{eqi1} \\& +\left.\! \left[k_a + \frac{i_{1,a}\cdot\Lambda_{\text{mod},\,j}}{\cos^2i_a} \right]\sin\Lambda_{\text{mod},\,j} \right)  
-   \left[k_a + \frac{i_{1,a}\cdot\Lambda_{\text{mod},\,j}}{\cos^2i_a} \right]\!\big|R_0-X_{\text{s},a}\big|e^{k_a\cdot\Lambda_{\text{mod},\,j}}  = 0,
\end{split}
\end{equation}
where $k_a(\Lambda_{{\text{mod},\,j}})\equiv \tan(i_a(\Lambda_{{\text{mod},\,j}}))  =
\tan (i_{0,a}+i_{1,a}\cdot\Lambda_{{\text{mod},\,j}})$. 
Equation~\eqref{eqi1} is similar to~\eqref{eqdist}, and all the difference is in the
multiplier that occurs due to the variability in the pitch angle.
 
We obtained the spiral segment parameters for young and old subsystems simultaneously
by minimizing the $\mathfrak{L}$ 
function~\eqref{fun} taking into account the new equation~\eqref{eqi1} for determining
$w$. We found that no segment has a significant value of $i_1$. Therefore,  this system
of segments can be described by spiral segments with the constant pitch angles.

\begin{figure}
   \centering
  \includegraphics[width=8cm, angle=0]{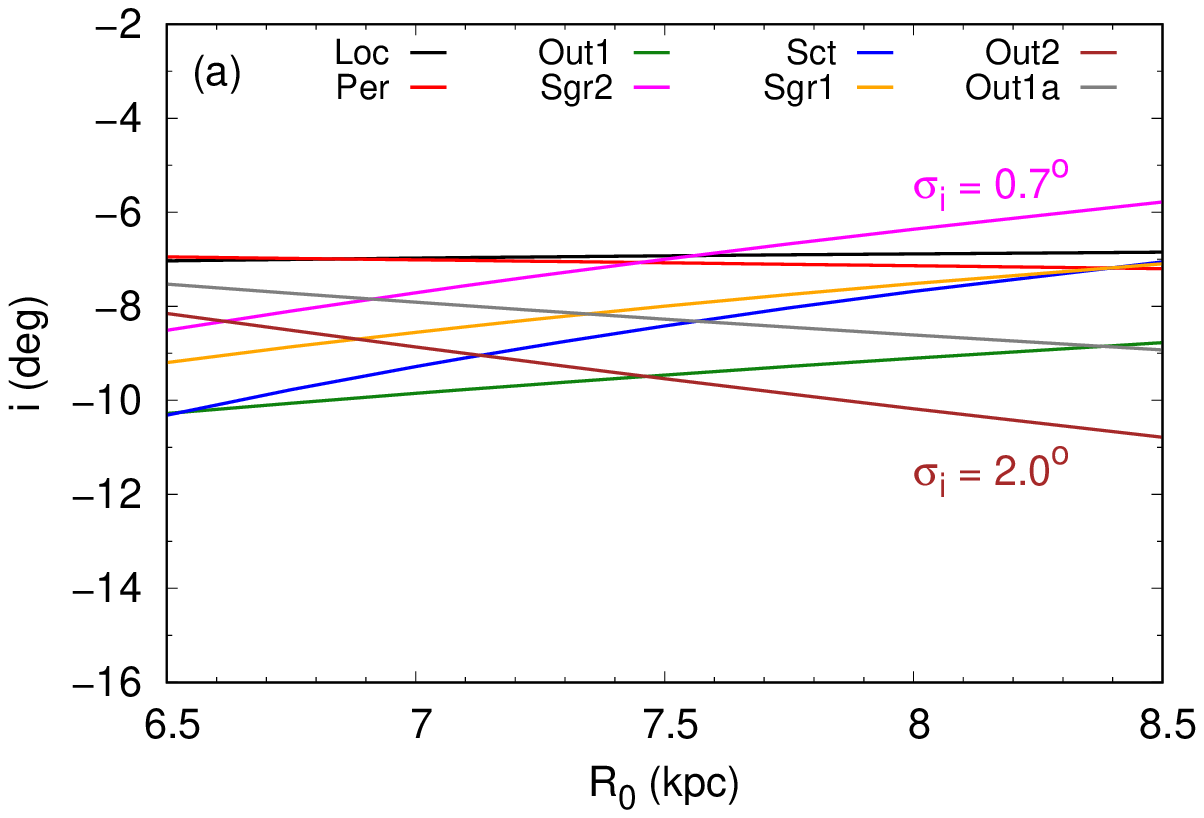}
  \includegraphics[width=8cm, angle=0]{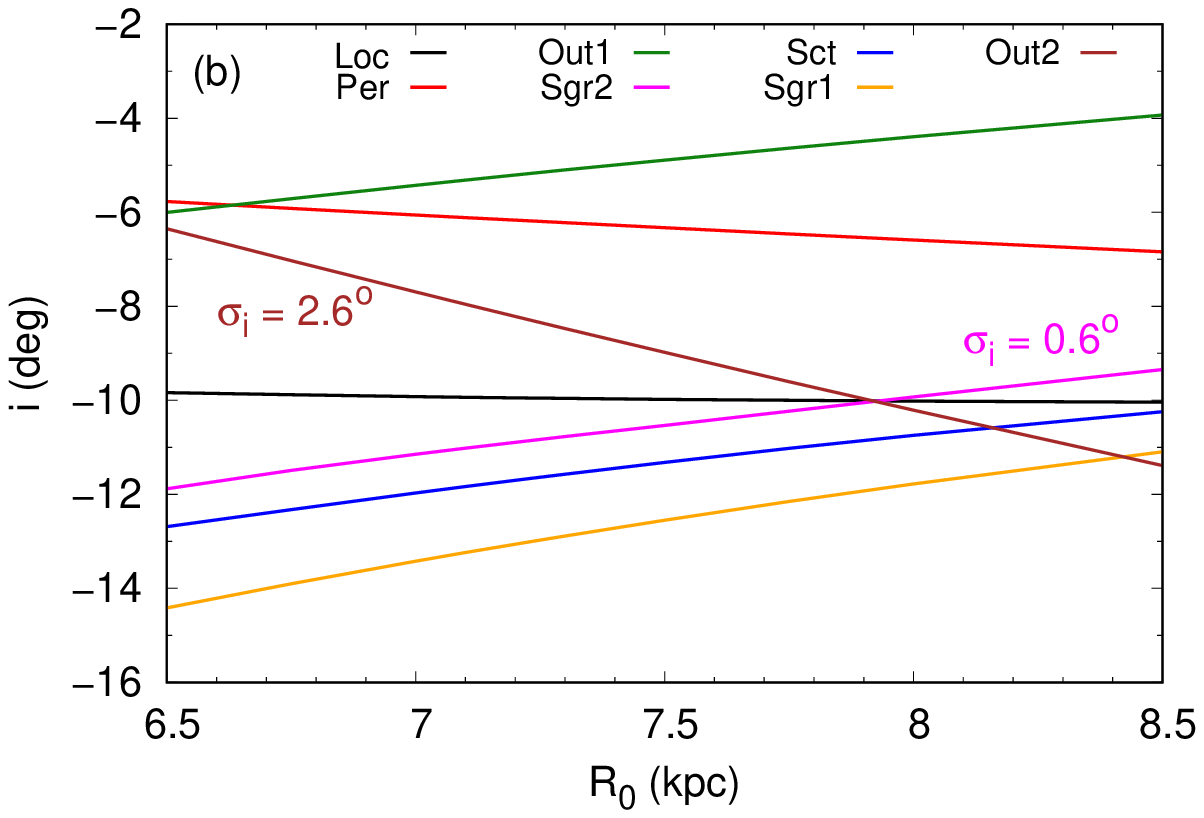}
   \caption{\rm\!Dependence of the pitch angles on $R_0$ for old (a)  and young (b) segments.
$\sigma_i$ denotes the characteristic value of the pitch angle uncertainty.} 
   \label{ii}
   \end{figure}

It was also interesting to investigate the dependence of the pitch angles on the
assumed value of $R_0$. We fixed the $R_0$ values and optimized the other parameters of
segments. Figure~\ref{ii} shows the dependence of the pitch angle on $R_0$ for young and
old segments. One can notice that although the spread of the pitch angle values for young
segments is greater than that for old ones, the slope of the dependency on $R_0$ is generally
the same for young and old segments. For several segments pitch angles change
significantly over the considered range of $R_0$ values. For example, the pitch angle of
both the young and old Sagittarius-2 segments changes by more than $2\fdg5$, while
$3\sigma_i$ does not exceed $2\fdg1$. On the contrary, the pitch angles of Outer-2
segments do not vary significantly because of the large uncertainty. Figure~\ref{ii} demonstrates that
in general the value $R_0$ can significantly affect the estimates of the pitch angles.

\begin{figure}
   \centering
  \includegraphics[width=8cm, angle=0]{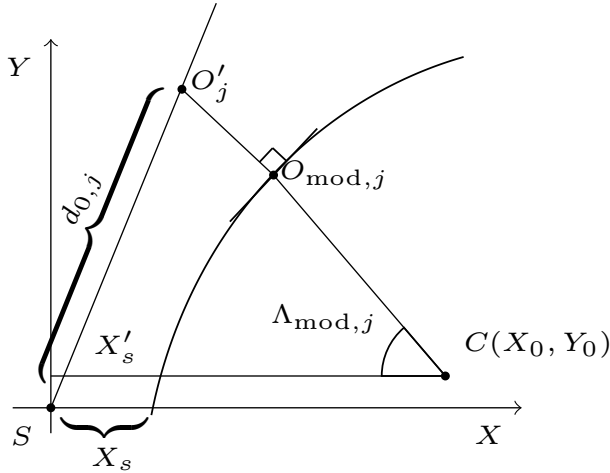}
   \caption{\rm\!Location of an object relative to the center line of the spiral segment
   with a pole offset from the line $SX$ of zero Galactic longitude
   (the nominal direction to the Galactic center).
   $C$ signifies the spiral pole, and $X_{\text{s}}'$ is the $X_{\text{s}}$ value corrected for the shift
   of~$C$. $O'_j$ represents the reduced position for the $j$-th object,
   $O_{\text{mod},j}$ is the point on a center line for which the distance $w$ from
   reduced position is minimal and ${\Lambda_{\text{mod},j}}$ is the Galactocentric longitude
   of the point $O_{\text{mod},j}$.} 
   \label{schemesh}
   \end{figure}

We also tried to clarify if  a bias exists in the pole of the spiral pattern from the
axis on which the Galactic longitude is equal to $0$. We consider the coordinates of a
pole to be $(X_0, Y_0)$ (see Fig.~\ref{schemesh}). Therefore, the solar
Galactocentric distance should be calculated as $R_0=\sqrt{X_0^2+Y_0^2}$. We also can
derive an angular bias of a pole:
$\theta = \arctan (Y_0/X_0)\,$.
   
As in the case of variable pitch angle, only the equation for determining the distance from
the reduced position to the center line of the segments changes significantly and
takes the form
\begin{equation}
\begin{split}
(X_{0,j} - X_0 )(\sin{\Lambda_{\text{mod},\,j}} - k\cos{\Lambda_{\text{mod},\,j}})  &- k|X_0-X'_{\text{s}}|e^{k{\Lambda_{\text{mod},\,j}}}  + \\
&+  (Y_{0,j}-Y_0)\left(k\sin{\Lambda_{\text{mod},\,j}} +  \cos{\Lambda_{\text{mod},\,j}} \right)  = 0,
\end{split}
\end{equation}
where $X_{\text{s}}'$ represents the coordinate of the intersection point of the segment's central line
with the line that is parallel to the $X$-axis and passes through the pole of the spiral pattern (see
Fig.~\ref{schemesh}). Simultaneous optimization of parameters for young and old segments 
yielded the following estimate of the pole coordinates: $X_0=7.22\pm0.19$\,kpc, $Y_0
=-0.06\pm0.15$\,kpc. Therefore, the bias of the pole is not significant and the angular
bias $\theta = -0\fdg5\pm 1\fdg3$ also can be treated as zeroth.

\section{Conclusions}
\label{sect:conclusions}
\noindent
We divided the sample of Cepheids into relatively young and old subsystems and
considered the spiral pattern consisting of 15 separate segments (seven traced by young
objects and eight traced by old ones). Estimates of the parameters were obtained jointly for
the two subsystems of segments following the maximum likelihood method for the spatial
modeling of spiral arm segments which takes into account the natural dispersion across the
segment and the uncertainty in the distance moduli. The obtained estimate \linebreak $R_0 =
7.23^{+0.19}_{-0.18}$\,kpc in the modern LMC calibration (de Grijs et al. 2014) corresponds to the value
$R_0={8.08^{+0.21}_{-0.20}}|_{\text{stat.}} {}^{+0.38}_{-0.36}|_{\text{cal.}}$\,kpc,
which is consistent with modern estimates such as $R_0=8.34 \pm 0.18$ kpc  obtained by Xu
et al. (2018) from kinematics of O stars and masers, $R_0=8.178 \pm 0.013\,(\text{stat.})
\pm 0.022\,(\text{sys.})$~kpc (GRAVITY Collaboration et al. 2019) and  $R_0=7.946 \pm
0.050\,({\text{stat}}) \pm 0.032\,({\text{sys}})$~kpc (Do et al. 2019) from modelling S0-2
orbit.

For Perseus and Sagittarius-2  segments a significant difference between the pitch angles
and $X_{\text{s}}$ values was found, and confidence regions for this segments also show the
difference between central lines. To interpret this displacement for Perseus and
Sagittarius-2 objects we carried out the dynamic modeling  of the motion of objects which
were formed in a spiral arm segment.  At  $\Omega_{\text{p}} = 25.2 \pm
0.5$\,km\,s$^{-1}$\,kpc$^{-1}$ (Dambis et al. 2015), the observed displacement  can be
explained by the value $u = 10 \pm1.5$\,km\,s$^{-1}$.

We demonstrated that spiral segments traced by Cepheids do not possess a significant gradient
in the pitch angle. Also the set of logarithmic segments do not need the pole of the
spirals to be displaced, so the location of the geometric pole of the spiral pattern is
consistent with the conventional direction to the Galactic center.

\section*{ACKNOWLEDGMENTS}
A.V.V.\ acknowledges support from the Russian Science Foundation, Grant No.~18-12-00050.
We are grateful for the helpful comments from the anonymous referee.

\section*{REFERENCES}

\begin{enumerate}

 \item   Berdnikov, L. N., Dambis, A. K., \& Vozyakova, O. V. 2000, AAS, 143, 211.   
 
 \item Bobylev, V.V., \& Bajkova, A.T. 2014, MNRAS, 437, 1549.
 
 \item  Braga, V. F., Bhardwaj, A., Contreras Ramos, R., Minniti, D., Bono, G., de Grijs,
 R., Minniti, J. H., \& Rejkuba, M. 2018, A\&A, 619, A51.
 
\item Casetti-Dinescu, D. I., Girard, T. M., J\'ilkov\'a, L., et al. 2013, AJ, 146, 33.
  
 \item  Do, T., Hees, A., Ghez, A., Martinez, G. D., Chu, D. S., et al. 2019, Science, 365, 664. 

  \item Dambis, A. K., Berdnikov, L. N., Efremov, Yu. N., Kniazev, A. Yu., Rastorguev, A.~S., Glushkova, E. V., Kravtsov, V. V., Turner, D. G., Majaess, D. J., \& Sefako, R. 2015, Astron. Lett., 41, 489.
  
  \item de Grijs, R., Wicker, J. E., \& Bono G. 2014, AJ, 147, 122. 
  
  \item Efremov, Yu. N. 2003, ARep, 47, 1000.  
  
  \item Efremov, Yu. N. 2011, ARep, 55, 108.  
  
  \item Englmaier, P., Pohl, M., \& Bissantz, N. 2011, MSAIS, 18, 199. 
  
  \item Francis C., \& Anderson, E. 2012, MNRAS, 422, 1283.
  
  \item 
  GRAVITY Collaboration (Abuter R. et al.),
  2019, \aap, 625, L10

  \item  Joshi, Y. C., \& Joshi, S. 2014, NewAst, 28, 27.  

  \item Karimova, D. K., \& Pavlovskaya, E. D. 1974, Soviet Astronomy, 17, 470. 
  
  \item
Kondrat'ev, A. S., \& Orlov, V. V. 2008, Astron. Lett., 34, 537.

 \item L\'epine, J. R. D., Mishurov, Yu. N., \& Dedikov, S. Yu. 2001, ApJ, 546, 234.

  \item Mel'nik A. M., Rautiainen P., Berdnikov L. N., Dambis A. K., \& Rastorguev A.S. 2015, Astron. Nachr., 336, 70.

  \item  Nikiforov, I., \& Veselova, A. 2015, BaltA, 24, 387 
  
  \item Nikiforov, I. I., \& Veselova, A. V. 2018a, Astron. Lett., 44, 81.
   
  \item Nikiforov, I. I., \& Veselova, A. V. 2018b, Astron. Lett., 44, 763. 
  
  \item Ninkovi\'c, S. 1992, Astron. Nachr., 313, 83.
  
  \item Savchenko, S. S., \& Reshetnikov, V. P.
  2013, MNRAS, 436, 1074.
     
  \item  Vall\'ee, J. P. 2008, AJ, 135, 1301.

  \item  Vall\'ee, J. P. 2015, MNRAS, 450, 4277.

  \item
  van de Hulst~H.C., Muller~C.A., \& Oort~J.H. Bull.
  Astron. Inst. Netherl., 1954, 12, 117.

  \item Veselova A. V., \& Nikiforov, I. I. 2018, in
  MSA Conf. Abstr. Vol. 1, eds. O. Yu. Malkov, A. S. Rastorguev, N.~N.~Samus' (Moscow:
  IZMIRAN), 100.
   
  \item Xu, Y., Hou, L.-G., \& Wu, Y.-W. 2018, RAA, 18, 146. 
   
\end{enumerate}


\end{document}